\documentclass[a4paper]{article}

\usepackage{INTERSPEECH2020}
\usepackage{color}
\usepackage{bbm}

\title{Unified Mandarin TTS Front-end Based on Distilled BERT Model }
\name{Yang Zhang, Liqun Deng, Yasheng Wang}
\address{
  Huawei Noah's Ark Lab
  }
\email{\{zhangyang86,dengliqun.deng,wangyasheng\}@huawei.com}

\begin{document}

\maketitle
\begin{abstract}
The front-end module in a typical Mandarin text-to-speech system (TTS) is composed of a long pipeline of text processing components, which requires extensive efforts to build and is prone to large accumulative model size and cascade errors.

In this paper, a pre-trained language model (PLM) based model is proposed to simultaneously tackle the two most important tasks in TTS front-end, i.e., prosodic structure prediction (PSP) and grapheme-to-phoneme (G2P) conversion. We use a pre-trained Chinese BERT\cite{devlin2019bert} as the text encoder and employ multi-task learning technique to adapt it to the two TTS front-end tasks. Then, the BERT encoder is distilled into a smaller model by employing a knowledge distillation technique called TinyBERT\cite{jiao2019tinybert}, making the whole model size $25\%$ of that of benchmark pipeline models while maintaining competitive performance on both tasks. With the proposed the methods, we are able to run the whole TTS front-end module in a light and unified manner, which is more friendly to deployment on mobile devices.

\end{abstract}
\noindent\textbf{Index Terms}: speech synthesis, TTS front-end, pre-trained language models, multi-task learning

\section{Introduction}

\begin{figure}[t]
	\centering
	\includegraphics[width=\linewidth]{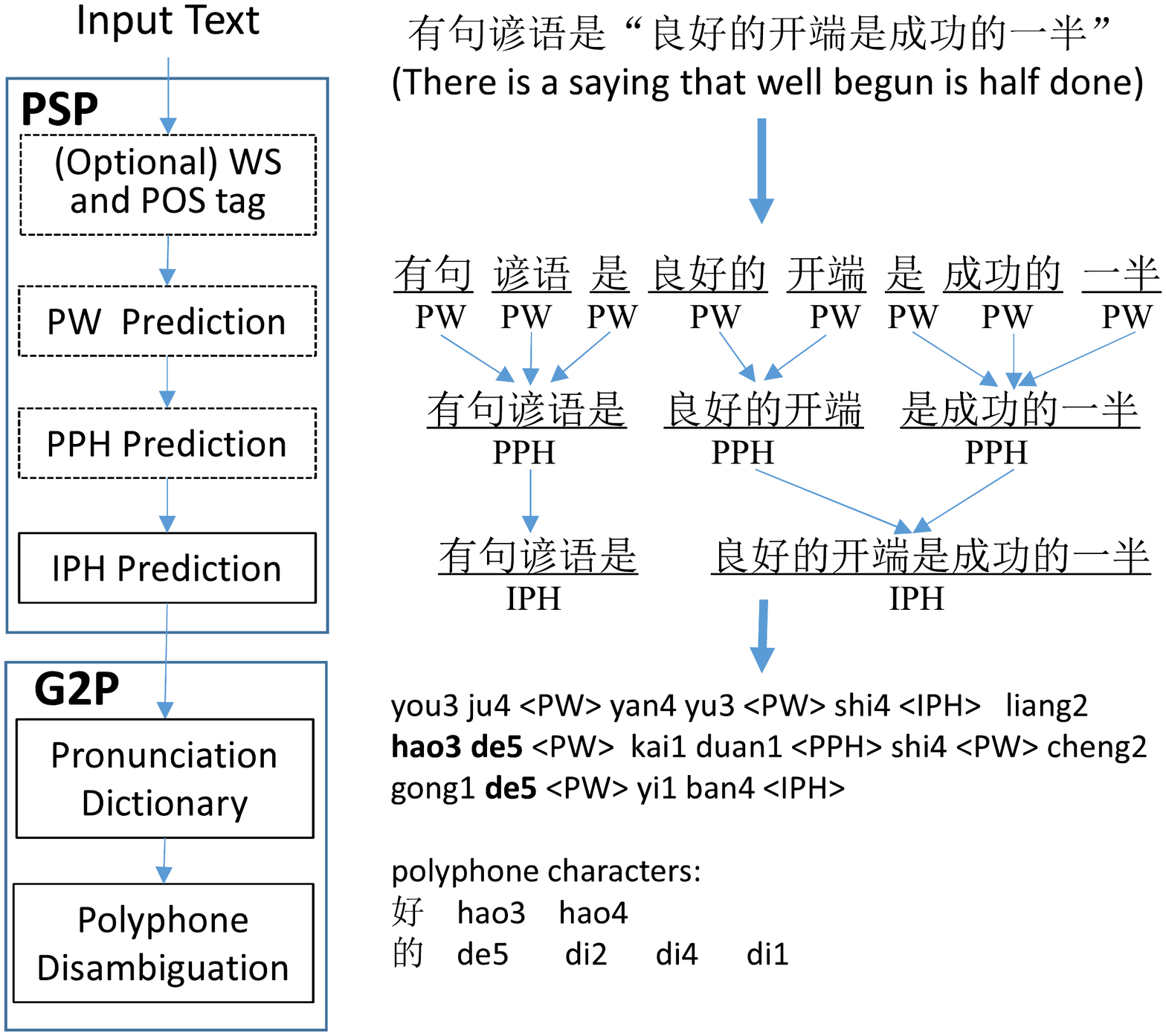}
	\caption{An example of Mandarin TTS front-end processing.}
	\label{fig:frontend}
\end{figure}

In Mandarin text-to-speech (TTS) synthesis, the front-end module has a strong impact on the intelligibility and naturalness of synthesized speech. As shown in Figure~\ref{fig:frontend}, a typical Mandarin TTS front-end module consists at least two main components, i.e., prosodic structure prediction (PSP) and grapheme-to-phoneme (G2P) conversion. PSP predicts the prosodic boundaries including prosodic word (PW), prosodic phrase (PPH) and intonational phrase (IPH), while G2P converts each Mandarin character into its pronunciation. Most existing TTS systems implement these two components or each processing step of them individually using either rule based models (e.g., syntactic trees based rules for PSP \cite{zhang2016mandarin}, dictionary matching et al based polyphone disambiguity \cite{huang2010disambiguation}), or statistical learning models (like CRF \cite{qian2010automatic} and LSTM/Attention based NN models \cite{pan2019mandarin,lu2019self} for PSP, \cite{cai2019polyphone,shan2016bi} for Mandarin G2P). The front-end module hence becomes a long pipeline of such individual steps as presented in Figure~\ref{fig:frontend}. As a result, it becomes a complicated and laborious work to build and maintain such a front-end, and also the storage and computation of the various front-end models hinder the deployment of TTS systems onto mobile devices. 

The aim of this work is to model PSP and G2P simultaneously in a unified manner and provide all the necessary prosodic and pronunciation information for the TTS backend (e.g., the acoustic model Tacotron~\cite{wang2017tacotron} and Wavenet vocoder \cite{oord2016wavenet}). This idea of simplifying TTS front-end has gained great attention recently. Deep Voice~\cite{arik2017deep} attempts to simplify the front-end pipeline by replacing all the components with deep neural networks. In \cite{yang2019pre}, large fixed pre-trained models are used to extract text representation for both PSP and G2P components, which achieves comparable performance while saving the efforts to elaborately prepare input features for each component individually. Pan et al \cite{pan2019unified} proposed a unified front-end structure that models PSP and G2P as a single sequence-to-sequence neural model in an auto-regressive way. Most of these exsiting works try to compact the front-end steps into a unified module, but still suffer from complex architecture or intensive computation, and can not be trained in a end-to-end fashion. Thus, a more simple and effective front-end model needs to be explored.

Inspired by the success of pre-trained language model (PLM) in TTS front-end related tasks \cite{zhu2019probing,sun2019knowledge,du2019prosodic,talman2019predicting}, we proposed a novel unified Mandarin TTS front-end model based on Chinese BERT, and further compressed it by utilizing a knowledge distillation technique named TinyBERT\cite{jiao2019tinybert}. The contribution of this work can be summarized into the following aspects:
\begin{itemize}
	\item We proposed a unified front-end model that employs BERT as text encoder and jointly models the PSP and G2P components in a multi-task learning framework\cite{ruder2017overview}. 
	
	\item We employed a two-stage distillation strategy \cite{jiao2019tinybert} to further distill the front-end model into a more compact one, which facilitate deployment on edge devices.
	
	\item Our approached models achieves state-of-the-art results on both PSP and G2P tasks. The distilled model can improve $0.31\%$ in polyphone disambiguation accuracy, $1.45\%$ and $0.52\%$ in F1 score of PPH and IPH prediction respectively, while having only $25\%$ size of baseline front-end models.   
\end{itemize}

\section{Proposed Method}

\begin{figure}[t]
	\centering
	\setlength{\abovecaptionskip}{0.cm}
	\includegraphics[width=0.8\linewidth]{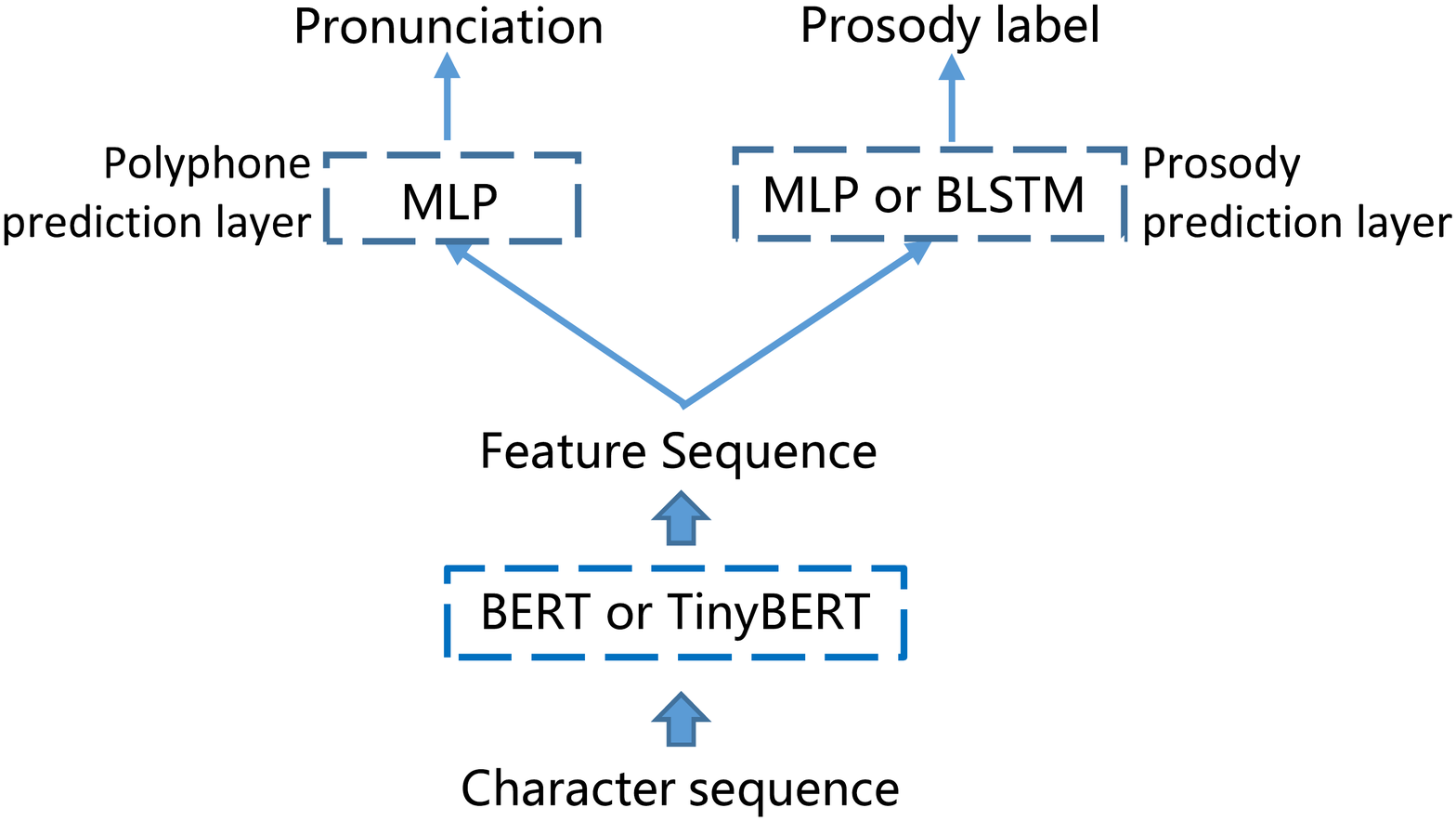}
	\caption{Architecture of the proposed model.}
	\label{fig:architecture}
\end{figure}

As shown in Figure~\ref{fig:architecture} , our model consists three parts, text encoder adopted from pre-trained Chinese BERT, prediction layer of polyphone disambiguation and prediction layer of prosodic structure. The whole model is finetuned with multi-task learning and then compressed by distillation.

\subsection{Pre-trained BERT for text representations}

The raw input text is first encoded with a well-prepared Chinese BERT (or TinyBERT for efficient computation purpose), which is composed of a stack of Transformer blocks and pre-trained with a huge amount of Mandarin text data. Benefit from this pre-training, the model is assumed to be capable of capturing rich contextual and semantic information of Mandarin language, and hence facilitates the downstream NLP tasks (i.e., PSP and G2P tasks in this work). 

We adopt character-based BERT for text representation. That is, for each character in the input text, a linguistic feature will be generated. Optionally, it is also feasible to use Chinese pre-trained language model that outputs word represention, but this way additionally requires word segmentation as a prerequisite step.

\subsection{Polyphone disambiguation}

The  Mandarin G2P problem can be divided into two categories, i.e., g2p of monophonic characters and g2p of polyphonic characters. The pronunciation of monophonic characters can be easily determined by a pronunciation dictionary, while g2p of polyphonic characters is a non-trivial problem as it is highly context sensitive. Thus, polyphone disambiguation is the core challenge in Chinese G2P.

Based on the rich contextual features from BERT, we propose to use multi-layer perceptron (MLP) with softmax as the pronunciation prediction layer. The task is taken as a classification problem here, with the pronunciations of all the polyphonic characters modeled together as different classes. For each polyphonic character in sentence, a probability distribution of all the  pronunciation classes ($\hat{y}=(y_{1},...,y_{c})$, where $y_{i}$ is the probability of pronunciation class $i$, and $c$ is the total number of pronunciations) will be yielded.

Cross entropy is employed as the training loss, and the loss of polyphonic characters is counted and then averaged for a training sentence as follows:
\begin{equation}
\mathcal{L}_{poly}=-\frac{1}{\left|W^{x}\right|}\sum_{\omega \in W^{x}}\sum_{c}\mathbbm{1}\{c=k_{\omega}\}\times \log y_{c}
\end{equation}
where $W_{x}$ is the set of indices of the polyphonic words in the training sentence $x$, $\mathbbm{1}$ is the indicator function, $k_{\omega}$ is the true label of character $\omega$.

In inference, the monophonic characters in a sentence are first identified and their pronunciations are directly determined with a dictionary, then the pronunciation $P_{\omega}$ of polyphonic character $\omega$ is predicted by the model as the maximal probability class  in the output distribution, i.e.,
\begin{equation}
P_{\omega}=\mathop{\arg\max}(y_{1},...,y_{c})
\end{equation}

\begin{figure}[t]
	\centering
	\includegraphics[width=\linewidth]{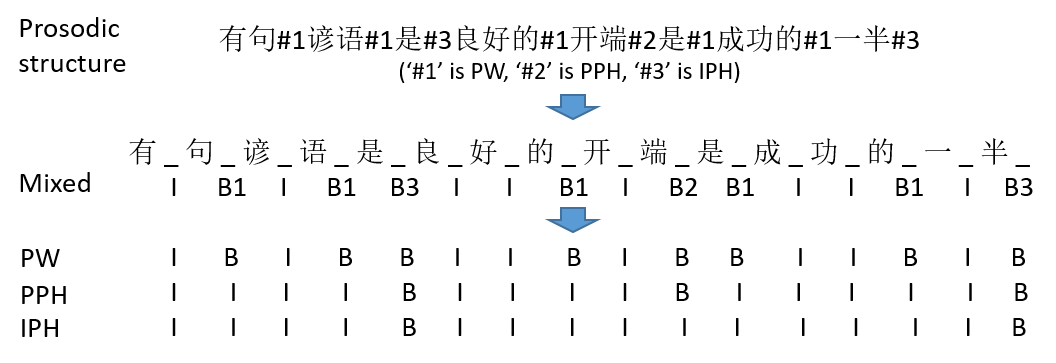}
	\caption{Example of mixed prosodic label. The three-level prosody label can be easily derived from the mixed result.}
	\label{fig:prosodiclabel}
\end{figure}

\subsection{Prosodic structure prediction}

We assume that there is essentially a prosodic break located behind each character and PSP is just to predict the class labels of these breaks. Four classes (labeled as {\textbf{I B1 B2 B3} for example) of prosodic break are considered, where label \textbf{I} stands for NO BREAK, and \textbf{B1, B2, B3} represents the prosodic break of prosodic word, prosodic phrase and intonational phrase respectively. Figure~\ref{fig:prosodiclabel} shows such a PSP example, where all the break labels can be directly predicted, instead of three prediction steps in traditional pipeline.
	
The prosody prediction layer is built with MLP or BLSTM network learned from the experience in \cite{zheng2018blstm,pan2019mandarin}. Similar to the polyphone prediction layer, the cross entropy loss is applied, but here the loss of all the characters are calculated.

\subsection{Multi-task learning}
The two output layers presented above are attached to the BERT encoder in parallel to form the unified front-end model, and a global loss is defined to train the model as follows, 
\begin{equation}
\mathcal{L}_{global}=\alpha_{1}\mathcal{L}_{poly}+\alpha_{2}\mathcal{L}_{prosody}
\end{equation}
where $\mathcal{L}_{poly}$ and $\mathcal{L}_{prosody}$ corresponds to the loss of polyphone prediction and prosody prediction respectively, while $\alpha_{1}$ and $\alpha_{2}$ are hyperparameters for loss balancing. 

On the other hand, it would be expensive to collect a training corpus both labelled pronunciation and prosodic boundary. Thus, we apply a mixed training strategy for training with two separate datasets. That is, the two datasets are blended in every training batch, and for each training sentence, only the loss of its labelled task is counted, while the other loss is set to zero. By this way, the weights of the whole model can be updated on both tasks continuously through the training process.

\subsection{TinyBERT distillation}
The original BERT model suffers intensive computations and large storage, which hindering its use to realtime applications and edge devices. To address this problem, knowledge distillation technique is most widely used, and TinyBERT\cite{jiao2019tinybert} is such a state-of-the-art distilled model, which is of the identical model architecture as BERT but with fewer Transformer layers and hidden units (see Section 3 for details).

During the distillation process, the original BERT model acts as the teacher model and TinyBERT as student. The distillation from BERT to TinyBERT can be divided into two parts, i.e., attention based distillation and hidden states based distillation (as showed in Fig.~\ref{fig:distillation}). For attention based distillation, the student tries to learn the multi-head attention matrix ($\mathbf{A^{S}}$) from that in the teacher network ($\mathbf{A^{T}}$). The learning objective is achieved by minimizing the following training loss ($\mathcal{L}_{attn}$),

\begin{figure}[t]
	\centering
	\includegraphics[width=\linewidth]{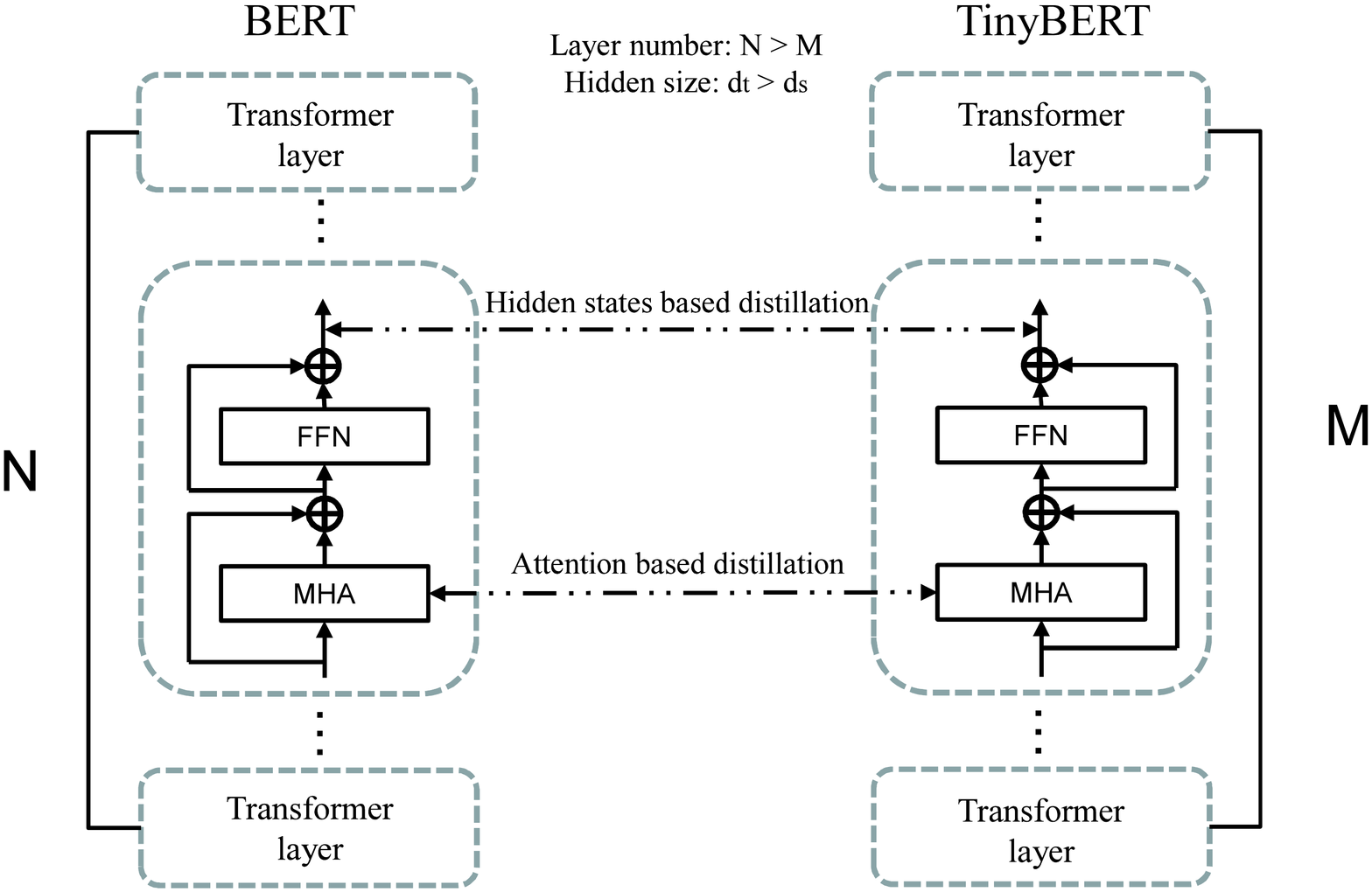}
	\caption{Illustration of TinyBERT distillation.}
	\label{fig:distillation}
\end{figure}
\begin{equation}
\mathcal{L}_{attn} = \frac{1}{h} \sum_{i=1}^{h}\operatorname{MSE} (\mathbf{A}_i^S, \mathbf{A}_i^T)
\end{equation}
where $h$ is the number of attention heads, and $\operatorname{MSE}$ means the mean squared error function.
Hidden states based distillation is to distill the knowledge in the output of Transformer layers. In each Transformer layer, it attempts to learn the reduced output state matrix $\mathbf{H^{S}}$ from the corresponding output in the teacher ($\mathbf{H^{T}}$). The training loss is defined as,
\begin{equation}
\mathcal{L}_{embed} = \operatorname{MSE} (\mathbf{H}^S\mathbf{W}_{h}, \mathbf{H}^T)
\end{equation}
where $\mathbf{H^{S}}$ and $\mathbf{H^{T}}$ is the output hidden states of a Transformer layer, $\mathbf{W_{h}}$ is a learnable linear projection which transforms $\mathbf{H^{S}}$ to the same dimension as $\mathbf{H^{T}}$.
The global distillation loss is finally calculated as the sum of $\mathcal{L}_{attn}$ and $\mathcal{L}_{embed}$ of all layers in the student model.

To enable TinyBERT efficient for TTS front-end tasks, it requires two-stage distillation, namely general distillation and task-specific distillation. The whole distillation work flow is illustrated in Figure~\ref{fig:diagram}, which includes the following steps:

\textbf{Step 1 General distillation}: Distilling a general TinyBERT model from the original pre-trained BERT model with the large-scale open domain data.

\textbf{Step 2 Finetune teacher model}: Taking BERT as the encoder of the front-end model and training the whole front-end with the TTS-specific training data (i.e., polyphone and PSP related training datasets). The BERT model will be finetuned during this training process.

\textbf{Step 3 Task distillation}: Taking the finetuned BERT model generated in Step 2 as the teacher and the general TinyBERT by Step 1 as the student, further distilling the knowledge of BERT to improve the student with the TTS-specific training data.

\textbf{Step 4 Finetune student model}: Replacing the encoder of the front-end model with the student TinyBERT model and repeating the training as Step 2.

\begin{figure}[t]
	\centering
	\includegraphics[width=\linewidth]{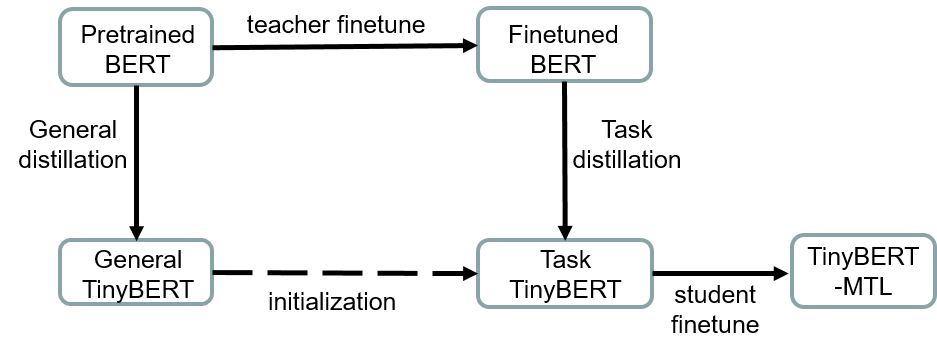}
	\caption{Flow diagram of the whole distillation algorithm.}
	\label{fig:diagram}
\end{figure}

\section{Experiments and discussion}
\subsection{Datasets and experiment settings}
An internal dataset of 340581 sentences and 661 polyphone characters in total is used for polyphone disambiguation, all the characters are labelled with pronunciation and each sentence has at least one polyphone character. We collected another corpus of 231964 sentences for prosodic structure prediction. Both datasets are divided into training set and test set with ratio 9:1. For automatic evaluation metrics, we report the result of character accuracy (ACC), sentence accuracy(SENT ACC) for polyphone disambiguation and  F1 score of PW, PPH, IPH for prosodic structure.

We use the original Chinese BERT base model in \cite{devlin2019bert} with 12 Transformer layers, 768 hidden state  to finetune the tasks. For distillation, we follow the previous work \cite{jiao2019tinybert} and use a student network (4 Transformer layers, 312 hidden states) to learn from the BERT base model. For the parameters of prediction layer, the MLP has 3 layers with hidden size of 512,  and the hidden units of BLSTM is 256.

\subsection{Systems built}
Systems for polyphone Disambiguation:

\textbf{WFST-based G2P}: Open source tool Phonetisaurus\cite{novak2016phonetisaurus} is used here, which is based on joint n-gram model in Weighted Finite-State Transducer (WFST) framework.

\textbf{BLSTM}: Following a recent method of neural network based G2P\cite{park2020g2pm}, the network consists one BLSTM layer with character embeddings and POS tag embeddings as input, followed with two fully connected layers and a softmax layer to predict pronunciation. 

\textbf{BERT-polyphone}: BERT encoder with only polyphone disambiguation task prediction layer.

Systems for prosodic structure prediction:

\textbf{CRF}: We implement CRF model with open source tool CRF++\footnote{https://taku910.github.io/crfpp/}. The used linguistic features include words, POS tags, and length of words etc. The 3-level prosody structure is modeled separately with three individual models, and the cascaded result is evaluated.

\textbf{BLSTM-CRF}: Following the settings of previous work\cite{pan2019mandarin}, embeddings of words, POS tags, word lengths, word distances are concatenated and fed into the multi-layer feedforward network(MFNN)-BLSTM-CRF network. Also, three individual models are built here for different prosody levels.

\textbf{BERT-prosody}: BERT encoder with only prosody structure prediction task prediction layer.

Systems for multi-task:

\textbf{BERT-MTL}: Proposed multi-task learning model with BERT encoder.

\textbf{TinyBERT-MTL}: Proposed multi-task learning model with tinyBERT encoder, which is distilled from BERT encoder.

\subsection{Results and analysis}

Table~\ref{tab:polyphone} shows the result of polyphone disambiguation. BERT-polyphone performs best, while BERT-MTL can achieve very close result with only $0.3$ accuracy lower. MTL model has to learn unrelated knowledges from both tasks, which may be the reason that MTL performs slightly worse than single task. TinyBERT-MTL beats the WFST-based G2P and BLSTM with improvement of $0.31\%$ in character accuracy and $0.62\%$ in sentence accuracy, while with a slightly accuracy lose compared to BERT-MTL.
\begin{table}[th]
	\caption{Accuracy of different systems in polyphone disambiguation}
	\label{tab:polyphone}
	\centering
	\begin{tabular}{ lll }
		\toprule
		\multicolumn{1}{c}{\textbf{Systems}} & 
		\multicolumn{1}{c}{\textbf{ACC}} & \multicolumn{1}{c}{\textbf{SENT ACC}}\\
		\midrule
		WFST-based G2P	& $96.48$ & $91.37$             \\
		BLSTM		& $93.78$  & $86.54$               \\
		BERT-polyphone	& $\mathbf{97.67}$  & $\mathbf{94.01}$      \\
		BERT-MTL		& $97.37$  & $93.71$              \\
		TinyBERT-MTL	& $96.79$  & $91.99$              \\
		\bottomrule
	\end{tabular}
	
\end{table}

Table~\ref{tab:prosody} presents the result of prosody structure prediction. The prosody break at the end of the sentence is not counted in F1 score, since it is just a trivial prediction. TinyBERT-MTL outperforms the baseline systems with nearly $2\%$ in all three F1 scores.  The CRF and BLSTM-CRF results are based on word segmentation predicted by Jieba\footnote{https://github.com/fxsjy/jieba} (a common used open source tool). 
\begin{table}[th]
	\caption{F1 score of different systems in prosody structure prediction}
	\label{tab:prosody}
	\centering
	\begin{tabular}{ llll }
		\toprule
		\multicolumn{1}{c}{\textbf{Systems}} & 
		\multicolumn{1}{c}{\textbf{PW-F1}} & \multicolumn{1}{c}{\textbf{PPH-F1}}
		& \multicolumn{1}{c}{\textbf{IPH-F1}}\\
		\midrule
		CRF	& $94.48$ & $71.44$     & $60.71$       \\
		BLSTM-CRF		& $94.23$  & $72.00$   & $65.33$           \\
		BERT-prosody	& $\mathbf{97.77}$  & $\mathbf{78.63}$  & $\mathbf{71.62}$   \\
		TinyBERT-prosody & $97.00$  & $76.93$  &   $70.22$         \\
		BERT-MTL		& $96.86$  & $78.10$  &   $70.99$         \\
		TinyBERT-MTL	& $96.40$  & $74.69$  &   $67.26$         \\
		\bottomrule
	\end{tabular}
	
\end{table}

Table \ref{tab:predictlayer1} demonstrates that MLP and BLSTM perform fairly close when using as prediction layer of PSP task in the model BERT-MTL. To model the transition between labels, previous work used CRF layer as the final output layer\cite{zheng2018blstm}. Here we compare the results of models with and without CRF output layer in the approached model in Table \ref{tab:predictlayer1}.  Similar  with the result of \cite{lu2019self}, slightly degrade is observed  when adding a CRF layer, which indicates that BERT encoder is more powerful to capture the label relationship in context than CRF.
\begin{table}[th]
	\caption{Analysis of prediction-layer of prosody structure}
	\label{tab:predictlayer1}
	\centering
	\begin{tabular}{ llll }
		\toprule
		\multicolumn{1}{c}{\textbf{prosody predicttion-layer}} & 
		\multicolumn{1}{c}{\textbf{PW-F1}} & \multicolumn{1}{c}{\textbf{PPH-F1}}
		& \multicolumn{1}{c}{\textbf{IPH-F1}}\\
		\midrule
		MLP	& $\mathbf{97.22}$ & $\mathbf{85.14}$     &  $86.71$       \\
		BLSTM		& $97.21$  & $85.10$   & $\mathbf{86.75}$           \\
		MLP-CRF	& $97.22$  & $84.53$  & $86.22$   \\
		BLSTM-CRF		& $97.21$  & $84.69$  &   $86.31$         \\
		\bottomrule
	\end{tabular}
	
\end{table}

As for polyphone task, recent study\cite{dai2019disambiguation} points  out that BLSTM is the best choice for prediction layer.  Our experiments gives a different answer that MLP has better accuracy (Table \ref{tab:predictlayer2}).  This result indicates that finetuned BERT encoder, in contrast to the frozen BERT in \cite{dai2019disambiguation}, can learn better polyphone related context knowledge than BLSTM.
\begin{table}[th]
	\caption{Analysis of prediction-layer of polyphone}
	\label{tab:predictlayer2}
	\centering
	\begin{tabular}{ lll }
		\toprule
		\multicolumn{1}{c}{\textbf{polyphone prediction-layer}} & 
		\multicolumn{1}{c}{\textbf{ACC}} & \multicolumn{1}{c}{\textbf{SENT ACC}}\\
		\midrule
		MLP	& $\mathbf{97.37}$  & $\mathbf{93.71}$             \\
		BLSTM		& $97.22$  & $93.41$               \\
		\bottomrule
	\end{tabular}
	
\end{table}

Table \ref{tab:modelsize} compares the model size of TinyBERT-MTL and benchmark front-end modules.  For benchmark, we choose the WFST-based G2P model, BLSTM-CRF based PSP model as they have better performance according to our experiments, and Jieba is used to do word segmentation and POS. We can see that the size of TinyBERT-MTL is about $25\%$ of benchmark, which is more feasible for mobile device deployment.
\begin{table}[th]
	\caption{Comparision of model size and inference time}
	\label{tab:modelsize}
	\centering
	\begin{tabular}{ lll }
		\toprule
		\multicolumn{1}{c}{\textbf{models}} & 
		\multicolumn{1}{c}{\textbf{model size}} & \multicolumn{1}{c}{\textbf{inference time}}\\
		\midrule
		Jieba(WS\&POS)	& $37$ MB  & less than $1$ ms             \\
		WFST-based(G2P)		& $143$ MB  & less than $1$ ms               \\
		BLSTM-CRF(PSP)		& $16$ MB  & $9$ ms              \\
		\midrule
		TinyBERT-MTL		& $47$ MB  & $21.8$ ms               \\
		\bottomrule
	\end{tabular}
	
\end{table}

The inference time of TinyBERT-MTL is $21.8$ ms on CPU server for a sentence of lengths $64$, this is averaged over $1000$ runs, and CPU info is Intel(R) Xeon(R) CPU E5-2690 v3 @ 2.60GHz. Although slower than benchmark models, it is fast enough since an increase of $20$ ms latency do not affect user experience in TTS.

\section{Conclusions}

In this paper, we proposed an unified end-to-end TTS front-end model for polyphone disambiguation and prosody structure prediction. The model accepts raw characters as input and outputs two predictions simultaneously, using BERT encoder and two task specific prediction layers. Furthermore, to make the model smaller for mobile deployment, we distilled BERT into a task specified TinyBERT. The final distilled and unified model outperforms the benchmark system in both tasks with only $25\%$ model size. In future work, we will investigate in acceleration of the model while optimizing the performance further close to the teacher model, and jointly learning of effective distillation and quantization might be a promise direction.

\bibliographystyle{IEEEtran}

\bibliography{mybib}

\begin{thebibliography}{10}
\providecommand{\url}[1]{#1}
\csname url@samestyle\endcsname
\providecommand{\newblock}{\relax}
\providecommand{\bibinfo}[2]{#2}
\providecommand{\BIBentrySTDinterwordspacing}{\spaceskip=0pt\relax}
\providecommand{\BIBentryALTinterwordstretchfactor}{4}
\providecommand{\BIBentryALTinterwordspacing}{\spaceskip=\fontdimen2\font plus
\BIBentryALTinterwordstretchfactor\fontdimen3\font minus
  \fontdimen4\font\relax}
\providecommand{\BIBforeignlanguage}[2]{{%
\expandafter\ifx\csname l@#1\endcsname\relax
\typeout{** WARNING: IEEEtran.bst: No hyphenation pattern has been}%
\typeout{** loaded for the language `#1'. Using the pattern for}%
\typeout{** the default language instead.}%
\else
\language=\csname l@#1\endcsname
\fi
#2}}
\providecommand{\BIBdecl}{\relax}
\BIBdecl

\bibitem{devlin2019bert}
J.~Devlin, M.-W. Chang, K.~Lee, and K.~Toutanova, ``Bert: Pre-training of deep
  bidirectional transformers for language understanding,'' in \emph{Proceedings
  of the 2019 Conference of the North American Chapter of the Association for
  Computational Linguistics: Human Language Technologies, Volume 1 (Long and
  Short Papers)}, 2019, pp. 4171--4186.

\bibitem{jiao2019tinybert}
X.~Jiao, Y.~Yin, L.~Shang, X.~Jiang, X.~Chen, L.~Li, F.~Wang, and Q.~Liu,
  ``Tinybert: Distilling bert for natural language understanding,'' \emph{arXiv
  preprint arXiv:1909.10351}, 2019.

\bibitem{zhang2016mandarin}
Z.~Zhang, F.~Wu, C.~Yang, M.~Dong, and F.~Zhou, ``Mandarin prosodic phrase
  prediction based on syntactic trees.'' in \emph{SSW}, 2016, pp. 160--165.

\bibitem{huang2010disambiguation}
F.-L. Huang, J.-H. Lin, and X.-W. Lin, ``Disambiguation for polyphones of
  chinese based on two-pass unified approach,'' in \emph{2010 International
  Computer Symposium (ICS2010)}.\hskip 1em plus 0.5em minus 0.4em\relax IEEE,
  2010, pp. 603--607.

\bibitem{qian2010automatic}
Y.~Qian, Z.~Wu, X.~Ma, and F.~Soong, ``Automatic prosody prediction and
  detection with conditional random field (crf) models,'' in \emph{2010 7th
  International Symposium on Chinese Spoken Language Processing}.\hskip 1em
  plus 0.5em minus 0.4em\relax IEEE, 2010, pp. 135--138.

\bibitem{pan2019mandarin}
H.~Pan, X.~Li, and Z.~Huang, ``A mandarin prosodic boundary prediction model
  based on multi-task learning,'' \emph{Proc. Interspeech 2019}, pp.
  4485--4488, 2019.

\bibitem{lu2019self}
C.~Lu, P.~Zhang, and Y.~Yan, ``Self-attention based prosodic boundary
  prediction for chinese speech synthesis,'' in \emph{ICASSP 2019-2019 IEEE
  International Conference on Acoustics, Speech and Signal Processing
  (ICASSP)}.\hskip 1em plus 0.5em minus 0.4em\relax IEEE, 2019, pp. 7035--7039.

\bibitem{cai2019polyphone}
Z.~Cai, Y.~Yang, C.~Zhang, X.~Qin, and M.~Li, ``Polyphone disambiguation for
  mandarin chinese using conditional neural network with multi-level embedding
  features,'' \emph{Proc. Interspeech 2019}, pp. 2110--2114, 2019.

\bibitem{shan2016bi}
C.~Shan, L.~Xie, and K.~Yao, ``A bi-directional lstm approach for polyphone
  disambiguation in mandarin chinese,'' in \emph{2016 10th International
  Symposium on Chinese Spoken Language Processing (ISCSLP)}.\hskip 1em plus
  0.5em minus 0.4em\relax IEEE, 2016, pp. 1--5.

\bibitem{wang2017tacotron}
Y.~Wang, R.~Skerry-Ryan, D.~Stanton, Y.~Wu, R.~J. Weiss, N.~Jaitly, Z.~Yang,
  Y.~Xiao, Z.~Chen, S.~Bengio \emph{et~al.}, ``Tacotron: Towards end-to-end
  speech synthesis,'' \emph{arXiv preprint arXiv:1703.10135}, 2017.

\bibitem{oord2016wavenet}
A.~v.~d. Oord, S.~Dieleman, H.~Zen, K.~Simonyan, O.~Vinyals, A.~Graves,
  N.~Kalchbrenner, A.~Senior, and K.~Kavukcuoglu, ``Wavenet: A generative model
  for raw audio,'' \emph{arXiv preprint arXiv:1609.03499}, 2016.

\bibitem{arik2017deep}
S.~{\"O}. Arik, M.~Chrzanowski, A.~Coates, G.~Diamos, A.~Gibiansky, Y.~Kang,
  X.~Li, J.~Miller, A.~Ng, J.~Raiman \emph{et~al.}, ``Deep voice: Real-time
  neural text-to-speech,'' in \emph{Proceedings of the 34th International
  Conference on Machine Learning-Volume 70}.\hskip 1em plus 0.5em minus
  0.4em\relax JMLR. org, 2017, pp. 195--204.

\bibitem{yang2019pre}
B.~Yang, J.~Zhong, and S.~Liu, ``Pre-trained text representations for improving
  front-end text processing in mandarin text-to-speech synthesis,'' \emph{Proc.
  Interspeech 2019}, pp. 4480--4484, 2019.

\bibitem{pan2019unified}
J.~Pan, X.~Yin, Z.~Zhang, S.~Liu, Y.~Zhang, Z.~Ma, and Y.~Wang, ``A unified
  sequence-to-sequence front-end model for mandarin text-to-speech synthesis,''
  \emph{arXiv preprint arXiv:1911.04111}, 2019.

\bibitem{zhu2019probing}
J.~Zhu, ``Probing the phonetic and phonological knowledge of tones in mandarin
  tts models,'' \emph{arXiv preprint arXiv:1912.10915}, 2019.

\bibitem{sun2019knowledge}
H.~Sun, X.~Tan, J.-W. Gan, S.~Zhao, D.~Han, H.~Liu, T.~Qin, and T.-Y. Liu,
  ``Knowledge distillation from bert in pre-training and fine-tuning for
  polyphone disambiguation,'' in \emph{2019 IEEE Automatic Speech Recognition
  and Understanding Workshop (ASRU)}.\hskip 1em plus 0.5em minus 0.4em\relax
  IEEE, 2019, pp. 168--175.

\bibitem{du2019prosodic}
Y.~Du, Z.~Wu, S.~Kang, D.~Su, D.~Yu, and H.~Meng, ``Prosodic structure
  prediction using deep self-attention neural network,'' in \emph{2019
  Asia-Pacific Signal and Information Processing Association Annual Summit and
  Conference (APSIPA ASC)}.\hskip 1em plus 0.5em minus 0.4em\relax IEEE, 2019,
  pp. 320--324.

\bibitem{talman2019predicting}
A.~Talman, A.~Suni, H.~Celikkanat, S.~Kakouros, J.~Tiedemann, and M.~Vainio,
  ``Predicting prosodic prominence from text with pre-trained contextualized
  word representations,'' \emph{arXiv preprint arXiv:1908.02262}, 2019.

\bibitem{ruder2017overview}
S.~Ruder, ``An overview of multi-task learning in deep neural networks,''
  \emph{arXiv preprint arXiv:1706.05098}, 2017.

\bibitem{zheng2018blstm}
Y.~Zheng, J.~Tao, Z.~Wen, and Y.~Li, ``Blstm-crf based end-to-end prosodic
  boundary prediction with context sensitive embeddings in a text-to-speech
  front-end,'' \emph{Proc. Interspeech 2018}, pp. 47--51, 2018.

\bibitem{novak2016phonetisaurus}
J.~R. Novak, N.~Minematsu, and K.~Hirose, ``Phonetisaurus: Exploring
  grapheme-to-phoneme conversion with joint n-gram models in the wfst
  framework,'' \emph{Natural Language Engineering}, vol.~22, no.~6, pp.
  907--938, 2016.

\bibitem{park2020g2pm}
K.~Park and S.~Lee, ``g2pm: A neural grapheme-to-phoneme conversion package for
  mandarinchinese based on a new open benchmark dataset,'' \emph{arXiv preprint
  arXiv:2004.03136}, 2020.

\bibitem{dai2019disambiguation}
D.~Dai, Z.~Wu, S.~Kang, X.~Wu, J.~Jia, D.~Su, D.~Yu, and H.~Meng,
  ``Disambiguation of chinese polyphones in an end-to-end framework with
  semantic features extracted by pre-trained bert,'' \emph{Proc. Interspeech
  2019}, pp. 2090--2094, 2019.

\end{thebibliography}


\end{document}